\documentclass[prx,twocolumn,amsmath,amssymb]{revtex4}

\usepackage{graphicx}
\usepackage{dcolumn}
\usepackage{bm}
\usepackage{mathrsfs}
\usepackage{subfigure}
\usepackage{natbib}
\usepackage{amsmath}
\usepackage{amssymb}
\usepackage{Jonasmacros}
\usepackage{mathptmx}
\usepackage{txfonts}
\usepackage{ifsym}
\usepackage[usenames,dvipsnames,svgnames]{xcolor}

\newcommand{\jonas}[1]{\textcolor{black}{#1}}

\begin{document}
\title{Shot noise as a probe of spin-\jonas{correlated} transport through single atoms}
\author{S. Pradhan}
\author{J. Fransson}
\affiliation{Department of Physics and Astronomy, Box 516, 75120, Uppsala University, Uppsala, Sweden}

\begin{abstract}
We address the shot noise in the tunneling current through a localized spin, pertaining to recent experiments on magnetic adatoms and single molecular magnets. We show that both \jonas{uncorrelated} and spin-\jonas{correlated} scattering processes contribute vitally to the noise spectrum. The spin-\jonas{correlated} scattering processes provide an additional contribution to the Landauer-B\"uttiker shot noise expression, accounting for correlations between the tunneling electrons and the localized spin moment. By calculating the Fano factor, we show that both super- and sub-Poissonian shot noise can be described within our approach. Our theory provides transparent insights to noise spectroscopy, consistent with recent experiments using local probing techniques on magnetic atoms.
\end{abstract}

\maketitle

\jonas{
\section{Introduction}
}
Noise spectroscopy is a versatile tool for studies of correlated matter. The signatures of the noise are directly related to internal fluctuations which, thereby, provide an immediate link to the excitations of the system.
Measurements of noise reveal, for instance, fractional charge of two-dimensional electron gas in the Hall regime \cite{fqhe1,fqhe2} and $2e$ charge of superconductor -- metal interface \cite{sc1,sc2,sc3}.
Detailed noise spectroscopy have, furthermore, revealed that many sources of instability of spin resonances used for qubit operations and, hence, the processes that dominates decoherence are amenable to improvements \cite{NatPhys.7.565,PhysRevLett.110.146804,NatNano.9.986}.

Particularly, investigations of spin noise through, e.g., atomic force microscopy opened for routes to use random fluctuations in spin ensembles to create spin order \cite{Science.307.408}. Moreover, optically probed noise spectroscopy has been utililized to record spontaneous spin noise associated with spin dynamics and magnetic resonance \cite{Nature.431.49}, electron and hole excitation spectra \cite{PhysRevLett.95.216603}, magneto-resonances due to electrons coupled to nuclear spins \cite{PhysRevLett.104.036601,PhysRevLett.108.186603,PhysRevB.89.081304}, and non-equilibrium spin noise \cite{PhysRevLett.113.156601}.
In magnetic set-ups such as quantum dots, or molecules with localized spin moment \cite{int1,int2,int3,int4}, noise spectroscopy opens routes to systematically investigate the underlying physics. Electrical current depends on the relative orientation of the localized spin and spin moment of the charge carrier. Hence, the charge transport couple with the spin dynamics and it will have the information about different energy scales of the spin system, all encoded in the noise \cite{item1,item2,item3,PhysRevLett.114.016602,PhysRevB.91.245404}.
Using charge transport for noise spectroscopy has theoretically been addressed for electrons coupled to molecular vibrations \cite{PhysRevB.67.165326,PhysRevB.70.205334,PhysRevB.90.075409} and local spin \cite{PhysRevB.68.085402}, while the major achievements have been made for optical probing techniques \cite{PhysRevB.88.201102,PhysRevB.92.180205,PhysRevB.93.033814}.

Recent shot noise measurements using scanning tunneling microscopy (STM) on single magnetic atoms, e.g., Fe and Co, adsorbed onto Au(111) surface showed a sub-Poissonian statistics \cite{PhysRevLett.114.016602}. Using a simple non-interacting Landauer-B\"uttiker picture for the shot noise, this sub-Poissonian signature was interpreted as evidence for spin-polarized transport, something which was further supported by density functional theory and linear conductance calculations. However, as both the STM tip as well as the substrate surface is non-spin-polarized, it is questionable whether a single (super-) paramagnetic spin moment would give rise to signatures of spin-polarization in the transport measurements. Previously reported experimental results have, on the contrary, provided strong evidences for non-spin-polarized transport properties in similar set-ups, see for instance \cite{Science.312.1021,Science.317.1199,Nature.467.1084,PhysRevLett.106.037205,PhysRevLett.111.157204,Science.350.417,PhysRevLett.114.106807}, both with and without spin orbit coupling in the metallic surface states.
\jonas{
Actually, all these measurements suggest the presence of correlations between the tunneling electrons and localized spin $\bfS$ through a coupling of the type $\sum_{\bfp\bfq}\psi^\dagger_\bfp\bfsigma\cdot\bfS\psi_\bfq+H.c.$, where $\psi_\bfq^\dagger$ ($\psi_\bfp$) denotes the creation (annihilation) spinor for electrons in the tip ($\bfq$) and substrate ($\bfq$). Since the Landauer-B\"uttiker picture does not sustain any explanation in terms of correlation effects at all, there is a calling for theoretical consistency between the conductance and noise measurements.
}

The lack of a transparent and consistent theoretical tool which enables simple and adequate analyses of the measurement data, justifies our reassessment of the theoretical description of shot noise in electron transport. While the non-interacting Landauer-B\"uttiker formulation is not applicable for the circumstances constituted by charge currents in presence of localized spin moments, it is one of few well-established approaches available.

\begin{figure}[t]
\includegraphics[width=\columnwidth]{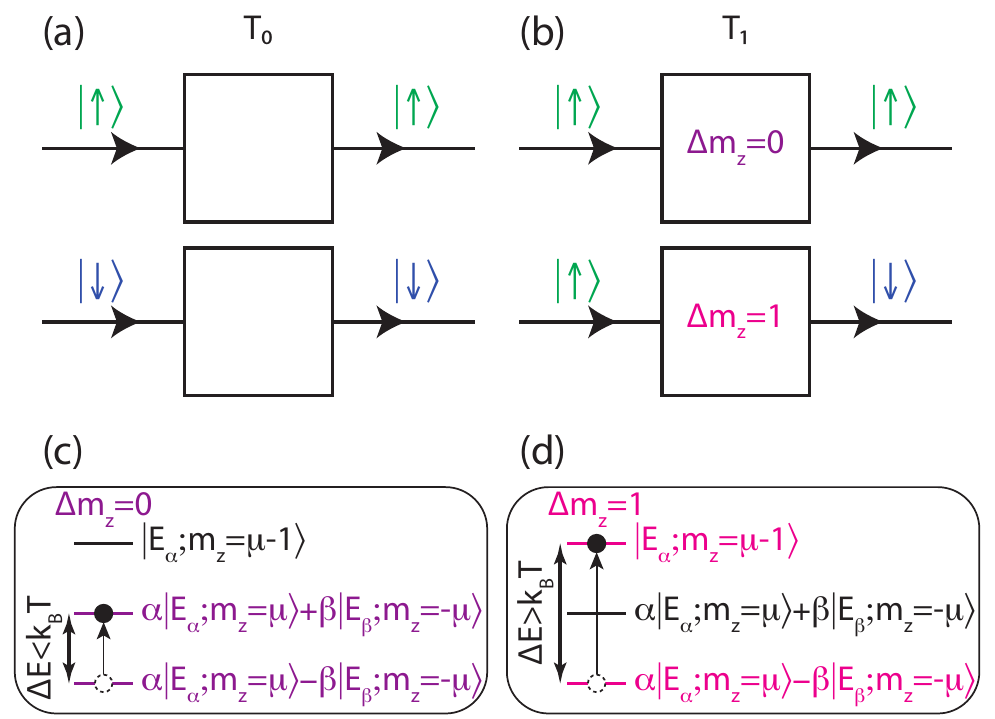}
 \caption{(Color online) (a), (b), Schematic description of the scattering processes involved in (a) the direct, rate $T_0$, and (b) indirect, rate $T_1$, tunneling. In the former, the spins constitute independent conduction channels whereas these channels are correlated through the internal structure in the latter. The internal structure is related to changes $\Delta m_z$ in the local spin angular momentum which is accompanied by conduction electron spin conserving (b), $\Delta m_z=0$, or spin-flip (b), $\Delta m_z=1$, processes. The changes $\Delta m_z$ arise from internal transitions (c) between different spin states in the scattering region. The corresponding energies for these transition may be activated either thermally, $\Delta E<k_BT$, or by external forces, $\Delta E>k_BT$.}
 \label{fig:scattering}
\end{figure}

In this article we address the problem of transport shot noise in presence of localized spin moments and derive a generalization of the Landauer-B\"uttiker theory. \jonas{The goal is to provide a simplified tool in the spirit of the Landauer-B\"uttiker formula for shot noise which, nonetheless, also includes correlations between the tunneling electrons and the localized spin.} We propose a model based on an interplay between direct and indirect tunneling, see Fig. \ref{fig:scattering}, where the direct tunneling electrons are unaffected by the localized spin while the indirect tunneling electrons undergo local exchange interactions with the spin. We show that this interplay precisely determines the characteristics of the shot noise. For a signal-to-noise ratio larger than one, a negligible contribution from the indirect tunneling leads to a Poissonian shot noise, in which limit our theory reduces to the Landauer-B\"uttiker picture. A stronger influence from the indirect tunneling\jonas{, increasing the ratio between the contributions from the indirect and direct tunneling,} leads to a sub-Poissonian shot noise. This is \jonas{in} agreement with the data reported in \cite{PhysRevLett.114.016602}. Oppositely, for small signal-to-noise ratios a large indirect tunneling contribution leads to a super-Poissonian character.

\jonas{
In Sec. \ref{sec-currentnoise} we derive the general expression for the current noise which comprises the spin fluctuations and we discuss the role of these on the transport measurements. Then, in Sec. \ref{sec-results} we discuss the results in terms of single magnetic moments and compare to recent measurements. The paper is discussed and summarized in Sec. \ref{sec-discussion}.
}

\jonas{
\section{Current noise}
\label{sec-currentnoise}
}

Here, we address the the quantum nature of the shot noise at low temperatures and consider non-equilibrium conditions. Fluctuations in the current can be characterized by calculating the Fourier transformation of the current-current correlation function $S(t,t')=\av{\anticom{\Delta I(t)}{\Delta I(t')}}/2=\av{\anticom{I(t)}{I(t')}}/2-\av{I(t)}\av{I(t'}$, where $\anticom{A}{B}=AB+BA$, whereas $\Delta I(t)=I(t)-\av{I(t)}$ denotes the deviation of the current $I(t)$ around its average value $\av{I(t)}$. The power spectrum of the noise is defined as the Fourier transform of $S(t,t')$, which for stationary conditions can be written
\begin{align}
S(\omega)=&
	\Bigl(
		S'(\omega,-\omega)
		-
		2
		\av{I}^2
	\Bigr)
	/2
	,
\label{ncorel}
\end{align}
where $S'(\omega,\omega') = \int\langle  I(t) I(t')\rangle e^{-i\omega t-i\omega't'} dtdt'$. The zero frequency ($\omega\rightarrow0$) noise is referred to as shot noise, which reduces to $S=2eI$ ($e$ -- electronic charge) in absence of electron correlations.

Our model comprises a localized spin moment $\bfS$ embedded in the tunnel junction between two normal metallic leads, referred to as the left ($L$) and right ($R$). The basic important assumption is that electrons can tunnel between the leads either by undergoing exchange interactions with the spin, with rate $T_1$, or not, with rate $T_0$, see. Fig \ref{fig:scattering}. The effective tunneling model is therefore formulated as ${\cal H}_T=\sum_{\bfp\bfq}\psi^\dagger_\bfp(T_0\sigma_0+T_1\bfsigma\cdot\bfS)\psi_\bfq+H.c.$. Here, $\psi_\bfk=(\cs{\bfk\up}\ \cs{\bfk\down})^t$ is the spinor for electrons in the left ($\bfk=\bfp$) or right ($\bfk=\bfq$) lead, wheras $\sigma_0$ and $\bfsigma$ are the identity matrix and vector of Pauli matrices, respectively. We notice that this tunneling model has been successfully employed previously in the description of, e.g., inelastic electron tunneling spectroscopy \cite{NanoLett.9.2414,PhysRevLett.102.256802,PhysRevLett.103.050801,PhysRevLett.103.176601}, as well as, electron spin resonance on single atomic spin using STM \cite{PhysRevB.66.195416,AdvPhys.61.117,SciRep.6.25584}.

\jonas{
We remark here that the introduced model for the tunneling is restricted to electron fluctuations around the electrochemical potentials $\mu_{L(R)}$ of the left (right) lead,  in that we assume constant rates for all electron tunneling processes. This is, however, justified since it is mainly the electrons around these chemical potentials that contribute to the transport properties of a junction between metals.
}

The overall model for the set-up is modelled by $\Hamil=\Hamil_L+\Hamil_R+\Hamil_S+\Hamil_T$. Here, $\Hamil_\chi=\sum_\bfk\psi^\dagger_\bfk(\dote{\bfk}-\mu_\chi)\psi_\bfk$ models the electrons in the lead $\chi=L,R$ with the energy dispersion $\dote{\bfk}$, assuming spin-degenerate electrons, relative to the chemical potential $\mu_\chi$. The voltage $V$ across the junction is defined by $eV=\mu_L-\mu_R$. The localized spin moment is modelled by $\Hamil_S$ for which the details are specified from case to case.

The current operator for the right lead is defined by
\begin{align}
I_R(t) =&
	-ie\sum_{\bfp\bfq}
		\psi^\dagger_\bfp(t)\hat\bfT(t)\psi_\bfq(t)
		-
		H.c.
		,
\end{align}
where we have introduced the notation $\hat\bfT=T_0\sigma_0+T_1\bfsigma\cdot\bfS$ and set $\hbar=1$. Considering the current-current auto-correlation function in Eq. (\ref{ncorel}), we use Wick's theorem to calculate each of the expectation values. In the stationary limit, it is justified to assume that the noises in the left and right leads are the equivalent. Hence, we calculate the contribution from the right lead, $S(t,t')=\av{I_R(t)I_R(t')}-\av{I_R(t)}\av{I_R(t')}$. As the disconnected diagrams of the first term exactly cancel the second term, we only need to consider the class of connected diagrams in the following. We write the auto-correlation function for the right lead\jonas{, to the second order in the tunneling rates,} as
\begin{widetext}
\begin{align}
S(t,t') =&
	-e^2
	\sum_{\stackrel{\scriptstyle \bfp\bfp'}{\sigma\sigma'}}
	\sum_{\stackrel{\scriptstyle \bfq\bfq'}{ss'}}
	\biggl(
		\av{\hat T_{\sigma s}(t)\hat T_{\sigma's'}(t')}
			F^>_{\bfq s\bfp'\sigma'}(t,t')
			F^<_{\bfq's'\bfp\sigma}(t',t)
		+
		\av{\hat T_{s\sigma}(t)\hat T_{s'\sigma'}(t')}
			F^>_{\bfp\sigma\bfq's'}(t,t')
			F^<_{\bfp'\sigma'\bfq s}(t',t)
\nonumber\\&\hspace{2cm}
		-
		\av{\hat T_{\sigma s}(t)\hat T_{s'\sigma'}(t')}
			G^>_{\bfq s\bfq's'}(t,t')
			G^<_{\bfp'\sigma'\bfp\sigma}(t',t)
		-
		\av{\hat T_{s\sigma}(t)\hat T_{\sigma's'}(t')}
			G^>_{\bfp\sigma\bfp'\sigma'}(t,t')
			G^<_{\bfq's'\bfq s}(t',t)
	\biggr)
	,
\label{eq-SRdef}
\end{align}
\end{widetext}
where the notation $F^>_{\bfq s\bfp\sigma}(t,t')=(-i)\av{\cs{\bfq s}(t)\cdagger{\bfp}(t')}$ and $F^<_{\bfq s\bfp\sigma}(t,t')=i\av{\cdagger{\bfp}(t')\cs{\bfq s}(t)}$ denote greater and lesser Green functions (GFs) for electron operators belonging to different leads, while $G^>_{\bfk\sigma\bfk\sigma'}(t,t')=(-i)\av{\cc{\bfk}(t)\csdagger{\bfk'\sigma'}(t')}$ and $G^<_{\bfk\sigma\bfk'\sigma'}(t,t')=i\av{\csdagger{\bfk'\sigma'}(t')\cc{\bfk}(t)}$ are used for electron operators within the same lead. 

\jonas{
Here, we have decoupled the two-electron propagators of the type
\begin{subequations}
\label{eq-twoelprop}
\begin{align}
\av{(\cdagger{\bfp}\cs{\bfq s})(t)(\csdagger{\bfp'\sigma'}\cs{\bfq's'}(t')}=&
	F^>_{\bfq s\bfp'\sigma'}(t,t')F^<_{\bfq's'\bfp\sigma}(t',t)
\label{eq-twoelpropFF}
,\\
\av{(\cdagger{\bfp}\cs{\bfq s})(t)(\csdagger{\bfq's'}\cs{\bfp'\sigma'}(t')}=&
	G^>_{\bfq s\bfq's'}(t,t')G^<_{\bfp'\sigma'\bfp\sigma}(t',t)
\nonumber\\&
	-
	F^>_{\bfp\sigma\bfq s}(t,t)F^>_{\bfp'\sigma'\bfq's'}(t',t')
	,
\label{eq-twoelpropGG}
\end{align}
\end{subequations}
where the disconnected diagram (second term) in Eq. (\ref{eq-twoelpropGG}) yields the contribution to the noise which cancel the product $\av{I_R(t)}\av{I_R(t')}$, as mentioned above.
}

The GFs $F^{</>}$ are expanded in terms of $G^{</>}$, using the Langreth's rules\jonas{, for instance,
\begin{align}
\bfF^{</>}_{\bfp\bfq}(t,t')=&
	\int
		\Bigl(
			\bfG^r_{\bfp\bfp'}(t,\tau)\hat{\bfT}(\tau)\bfG^{</>}_{\bfq'\bfq}(\tau,t')
\nonumber\\&\hspace{1cm}
			+
			\bfG^{</>}_{\bfp\bfp'}(t,\tau)\hat{\bfT}(\tau)\bfG^a_{\bfq'\bfq}(\tau,t')
		\Bigr)
	d\tau
	,
\end{align}
where the superscript $r/a$ denote the corresponding retarded/advanced GFs, whereas the bold face notation denotes matrices in spin 1/2 space.
}

Assuming that scattering between different states within the same lead is negligible, setting $G^{</>}_{\bfk\sigma\bfk\sigma'}=\delta_{\sigma\sigma'}\delta(\bfk-\bfk')G_{\bfk\sigma}^{</>}$, we obtain a closed formula for $S$. For leads in local equilibrium we can write $G_{\bfk\sigma}^{</>}(t,t')=(\pm i)f_\chi(\pm\omega)\exp\{-i\dote{\bfk}(t-t')\}$, where $f_\chi(\omega)=f(\omega-\mu_\chi)$ is the Fermi function.

\jonas{
\subsection{The role of spin fluctuations}
\label{ssec-spinnoise}
}
The correlation function $\av{\hat\bfT(t)\hat\bfT(t')}=T_0^2\sigma_0+T_1^2\bfsigma\cdot\bfchi(t,t')\cdot\bfsigma$ contains the direct and indirect tunneling processes, of which the latter depends on the spin fluctuations comprised in the spin-spin correlation function $\bfchi(t,t')=\av{\bfS(t)\bfS(t')}$.
\jonas{
As we are interested in the effects of the spin transitions on the properties of the shot noise, we neglect any life-time effects of the spin states and consider the local spin moment in the atomic limit.
Hence, e}xpanding $\bfchi$ in terms of the eigenstates and eigenenergies $\{\ket{a},E_a\}$ of $\Hamil_S$ we obtain
\begin{align}
\bfsigma\cdot\bfchi(t,t')\cdot\bfsigma=&
	\sum_{ab}
		\Big(2\chi^z_{ab}+\chi^{-+}_{ab} + \chi^{+-}_{ab} \Big)
		e^{iE_{ab}(t-t')}
\end{align}
where $\chi_{ab}^{z/-+/+-}=\bra{a}S^{z/-/+}\ket{b}\bra{b}S^{z/+/-}\ket{a}P(E_a)[1-P(E_b)]$ and $P(E_a)$ is an occupation factor for the state $\ket{a}$, whereas $E_{ab}=E_a-E_b$ is the energy associated with the transition.
\jonas{
In this way we incorporate the quantum nature of the localized spin moment, which is necessary in order to appropriately account for the role of the spin fluctuations on the shot noise.
}

The partitioning of $\bfchi$ into longitudinal $\chi^z$ and transverse $\chi^{\pm\mp}$ components reflects the differences in the allowed spin transitions, with respect to changes in the spin angular momentum $\Delta m_z$. The former transitions ($\chi^z$) do not involve any changes in the local spin angular momentum ($\Delta m_z=0$), Fig. \ref{fig:scattering} (c), and are accompanied by spin-conserving tunneling electrons, Fig. \ref{fig:scattering} (b) (upper). The latter ($\chi^{\pm\mp}$) concern unit changes in the local spin angular momentum ($|\Delta m_z|=1$), Fig. \ref{fig:scattering} (d), requiring spin-flip processes by the tunneling electrons, Fig. \ref{fig:scattering} (b) (lower).

The auto-correlation function given in Eq. (\ref{eq-SRdef}) contains terms involving different orders of $T_0$ and $T_1$, which are systematically collected, such that we rewrite the noise as $S(V) =2e^2n_Rn_L \sum_{nm}S_{nm}(V)$, where $S_{nm}(V) \propto T_0^nT_1^m$, and $n_\chi$ denotes the density of electron states in the lead $\chi$. It is then straightforward to see that the first two terms of Eq. (\ref{eq-SRdef}) give the highest order contributions in $T_0$ and $T_1$, that is, $S_{40}$, $S_{22}$, and $S_{04}$. By integrating out the time variables and assuming wide band metals in the leads, see for instance Ref. \cite{PhysRevB.81.115454}, the contribution proportional to $T_0^4$ can be written
\begin{align}
 S_{40}=&
	T_0^4
	\biggl(2k_BT-eV\coth\frac{eV}{2k_BT}\biggr)
	,
\end{align}
which in the low temperature limit ($k_BT/eV\ll1$) becomes $S_{40}\simeq  -T_0^4|eV|$. Similarly, the other contributions can written as $S_{22}=-k_BTT_0^2T_1^2\sum_{ab}\chi_{ab}$ and the $S_{04}=-k_BTT_1^4\sum_{ab}\chi^2_{ab}$ where we have defined $\chi_{ab}=2\chi_{ab}^z+\chi_{ab}^{-+}+\chi_{ab}^{+-}$.

Next, we consider the last two term of the Eq. (\ref{eq-SRdef}), which are quadratic in the tunneling rates, given by $S_{02}$ and $S_{20}$. Again, in the zero frequency limit we obtain
\begin{align}
S_{20}=&
	T_0^2eV\coth\frac{eV}{2k_BT}
	,
\end{align}
which tends to $S_{20}\simeq T_0^2|eV|$ for low temperatures, while
\begin{align}
S_{02}=&
	T_1^2\sum_{ab}\chi_{ab}(eV+E_{ab})\coth\frac{eV+E_{ab}}{2k_BT}
	.
\end{align}
Collecting all the terms leads to the total shot noise
\begin{align}
S=&
	2e^2
	n_Rn_L
	\Biggl\{
		k_BT
		\biggl[
			2T_0^4-T_1^2\sum_{ab}\Bigl(T_0^2+T_1^2\chi_{ab}\Bigr)\chi_{ab}
		\biggr]
\nonumber\\&
		+
		eVT_0^2(1-T_0^2)\coth\frac{eV}{2k_BT}
\nonumber\\&
		+
		T_1^2\sum_{ab}\chi_{ab}\Bigl(eV+E_{ab}\Bigr)
			\coth\frac{eV+E_{ab}}{2k_BT}
	\Biggr\}
	.
\label{eq-S}
\end{align}

\begin{figure}[t]
\includegraphics[width=\columnwidth]{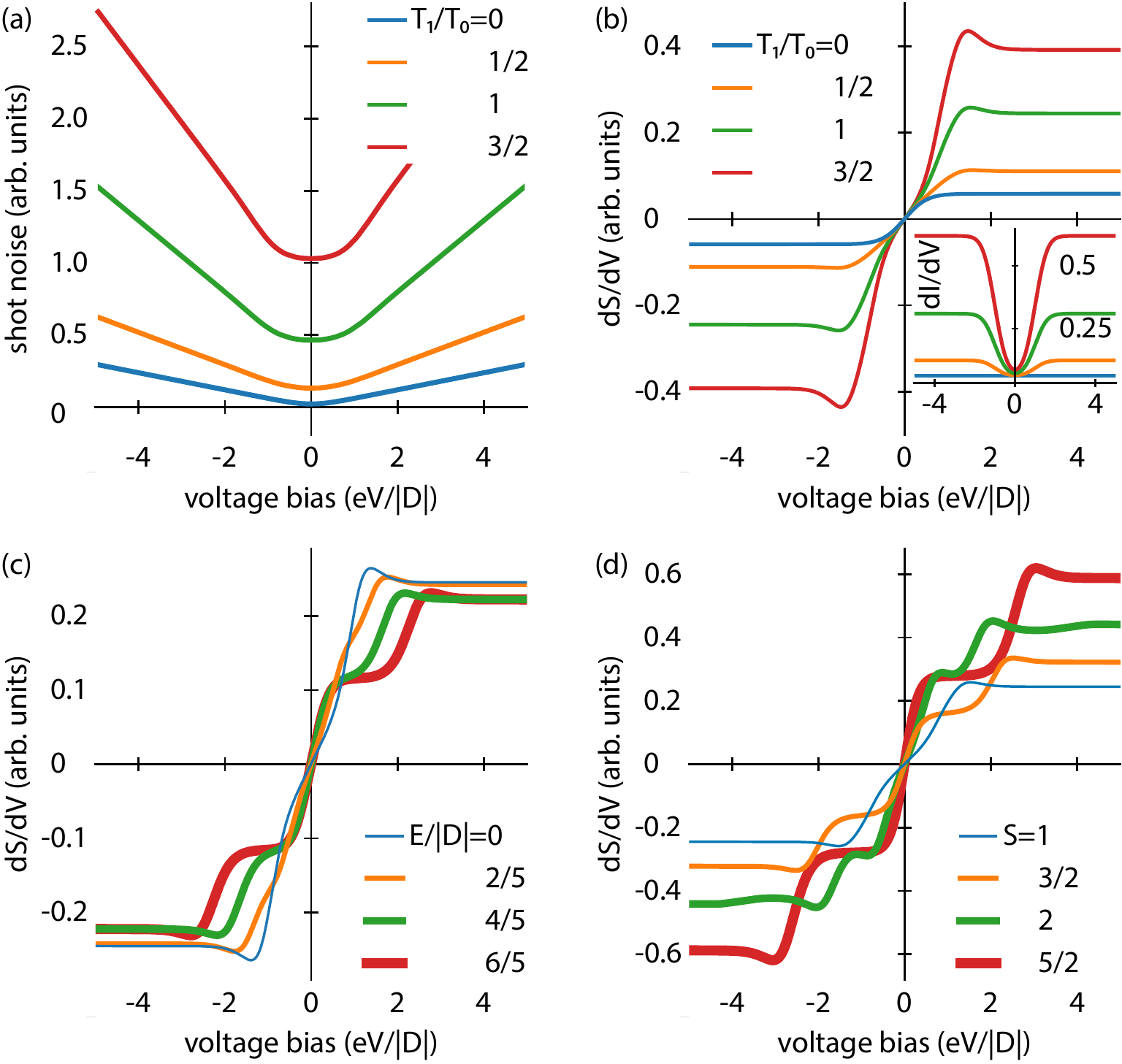}
 \caption{(Color online) Shot noise (a) and corresponding differential shot noise ($dS/dV$) (b) for varying ratio $T_1/T_0=0$, $1/2$, $1$, and $3/2$ as function of the voltage bias. Here, the spin $S=1$ and temperature $T=1$ K, whereas uniaxial and transverse anisotropies $D=1$ meV and $E/|D|=1/5$, respectively. The inset in panel (b) shows the corresponding differential conductance ($dI/dV$). (c) $dS/dV$ for varying $E/|D|=0$, $2/5$, $4/5$, and $6/5$ for spin $S=1$, and (d) for varying spin $S=1$, $3/2$, $2$, and $5/2$, for $T_1/T_0=1$ and $E/|D|=1/5$.}
 \label{fig:noise}
\end{figure}

First, it should be noticed that this expression is proportional to $2T_0^4k_BT+eVT_0^2(1-T_0)^2\coth{eV/2k_BT}$ in absence of the indirect tunneling processes ($T_1=0$), which in the low temperature regime ($k_BT\ll eV$) reduces to
\begin{align}
S=&
	2e^2n_Rn_Le|V|
		T_0^2(1-T_0^2)
	.
\end{align}
We, thus, retain the Landauer-B\"uttiker formula for the shot noise \cite{buttiker} in the limit of non-interacting tunneling electrons. We note that the above expression is easily generalized to include also spin-polarized leads. For a perfectly transmitting channel, the current fluctuations (noise) are expected to have less significance, which clear when the transmission $T_0$ tends towards unity, completely suppressing $S$. For weak transmission, on the other hand, we recover the full shot noise formula $S=2eI$. The corresponding differential noise calculated from Eq. (\ref{eq-S}) at $T_1=0$ is shown in Fig. \ref{fig:noise} (a), (b) (blue), which indicates a linear voltage dependence of the noise except near equilibrium, in agreement with previous results.
Signatures in the noise emerging from the exchange interactions between the tunneling electrons and localized spin require a finite rate $T_1$, to which the remainder of the article is devoted.

\jonas{
\section{Results}
\label{sec-results}
}

We make connection to the experiment in, e.g., Ref. \cite{PhysRevLett.114.016602} by using $H_S = -g\mu_B \mathbf{B}\cdot\mathbf{S}+DS_z^2+E(S_{+}^2+S_{-}^2 )/2$ to model the local spin structure. Here, the parameters $D$ and $E$ account for the uniaxial and transverse anisotropies, respectively, whereas $\bfB$ is an external magnetic field, $g$ is the gyromagnetic ratio, and $\mu_B$ is the Bohr magneton.

Using the result derived in Eq. (\ref{eq-S}), we investigate the influence of the local spin on the noise as function of the voltage, the excess noise. In Fig. (\ref{fig:noise}) (a), (b), we plot the shot noise $S$ and corresponding differential shot noise $dS/dV$, respectively, for increasing ratio $T_1/T_0$. The emergence of the dip and peak symmetrically located on either side of equilibrium signify the inelastic spin transition of the local spin which is assisted by the exchange of energy and spin angular momentum with the tunneling electrons. The increasing intensity of these features is consistent with the differential conductance of single paramagnetic moments using STM \cite{Science.312.1021,Science.317.1199,Nature.467.1084,PhysRevLett.106.037205,PhysRevLett.111.157204,Science.350.417,PhysRevLett.114.106807}, which can be seen in the inset.
For a finite but small transverse anisotropy $E<k_BT$, transitions that do not require the exchange of spin angular momentum are thermally activated, see Fig. \ref{fig:scattering} (c), which leads to a increased equilibrium shot noise, Fig. \ref{fig:noise} (a).
Increasing $E$ increases the energy split between the excited states for integer spins, for which the dip/peak is expected to shift towards higher voltages. This can be seen in Fig. \ref{fig:noise} (c), where we plot $dS/dV$ for increasing $E$.
Higher spins with more excitations are expected to reveal more features in the differential excess noise, which is verified in Fig. \ref{fig:noise} (d), where we display results for spin $S=1$, $3/2$, $2$, and $5/2$.

\begin{figure}[t]
\includegraphics[width=\columnwidth]{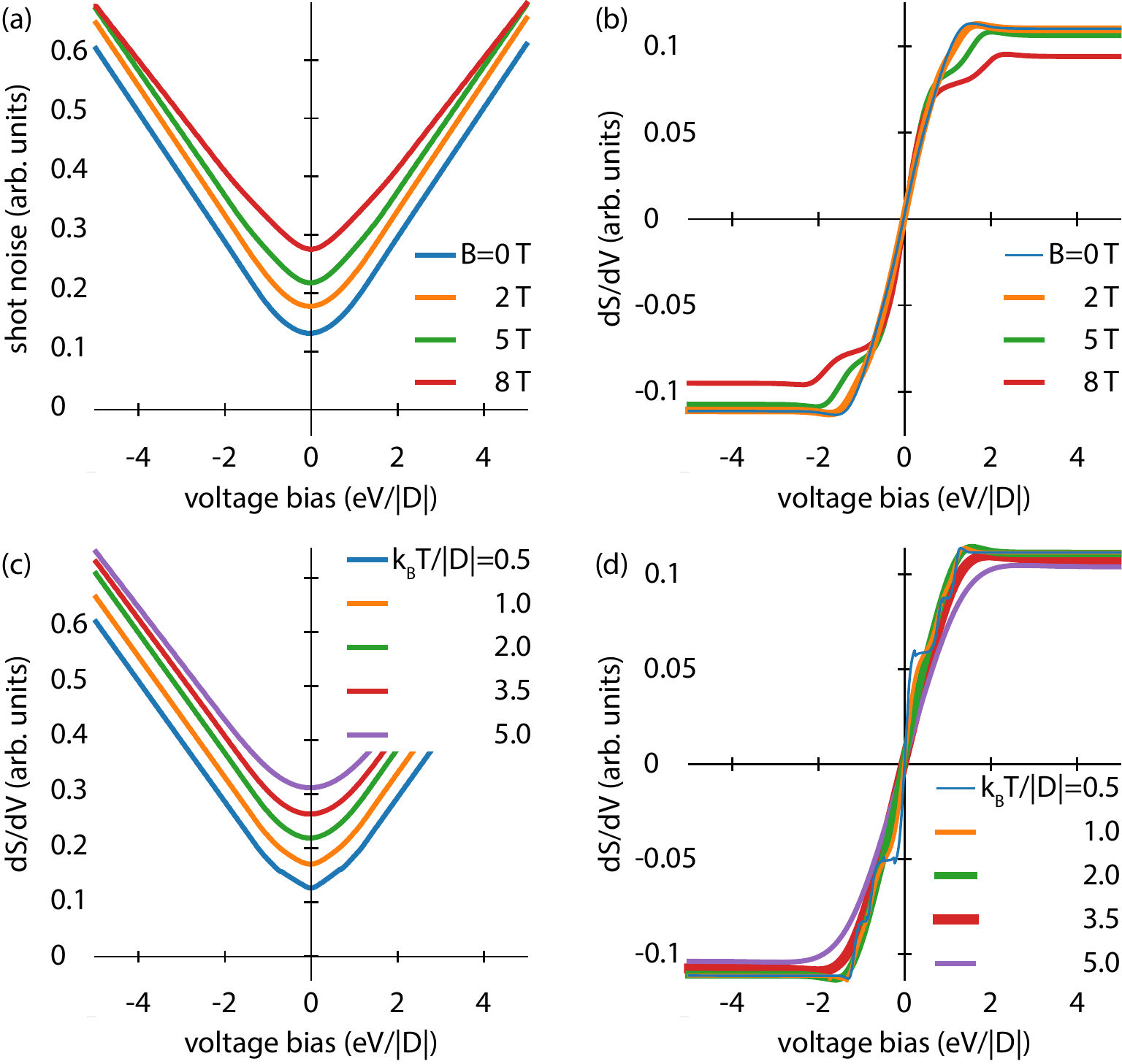}
 \caption{
 \jonas{
 (Color online) Shot noise (a), (c), and corresponding $dS/dV$ (b), (d), for varying external magnetic field $\bfB=B\hat{\bf z}$, with $B=0$, 2, 5, and 8 T (a), (b), and temperature $k_BT/|D|=0.5$, 1, 2, 3.5, and 5 (c), (d), as function of the voltage bias. Here, the spin $S=1$, $T_1/T_0=1/2$, and $E/|D|=0.2$, and in panels (a), (b),  $T=2$ K. The curves in panels (a) and (c) are off-set by $0.05n$, $n$ non-negative integer, for clarity.}
}
\label{fig:BTnoise}
\end{figure}

\jonas{
The impact of externally applied magnetic fields ${\bf B}=B\hat{\bf z}$ is plotted in Fig. \ref{fig:BTnoise} (a), (b), showing the shot noise and corresponding differential shot noise as function of the voltage bias for increasing magnetic field strengths $B=0$, 2, 5, and 8 T. As the Zeeman split of the local spin increases with the magnetic field, the correlated noise which is associated with the spin transitions ($\bfchi$) are suppressed at low voltages and are only accessed through the energy disposal at higher voltages. In the shot noise this is illustrated by that the low voltage characteristics goes from a rounded $U$ shape to a more chevron like appearance, with increasing magnetic field strengths. The corresponding features in the differential shot noise is the emerging step like characteristics with the voltage for increasing magnetic field strengths. These steps reflect the increased energy spacing in the spin excitation spectrum.
}

\begin{figure}[t]
\includegraphics[width=\columnwidth]{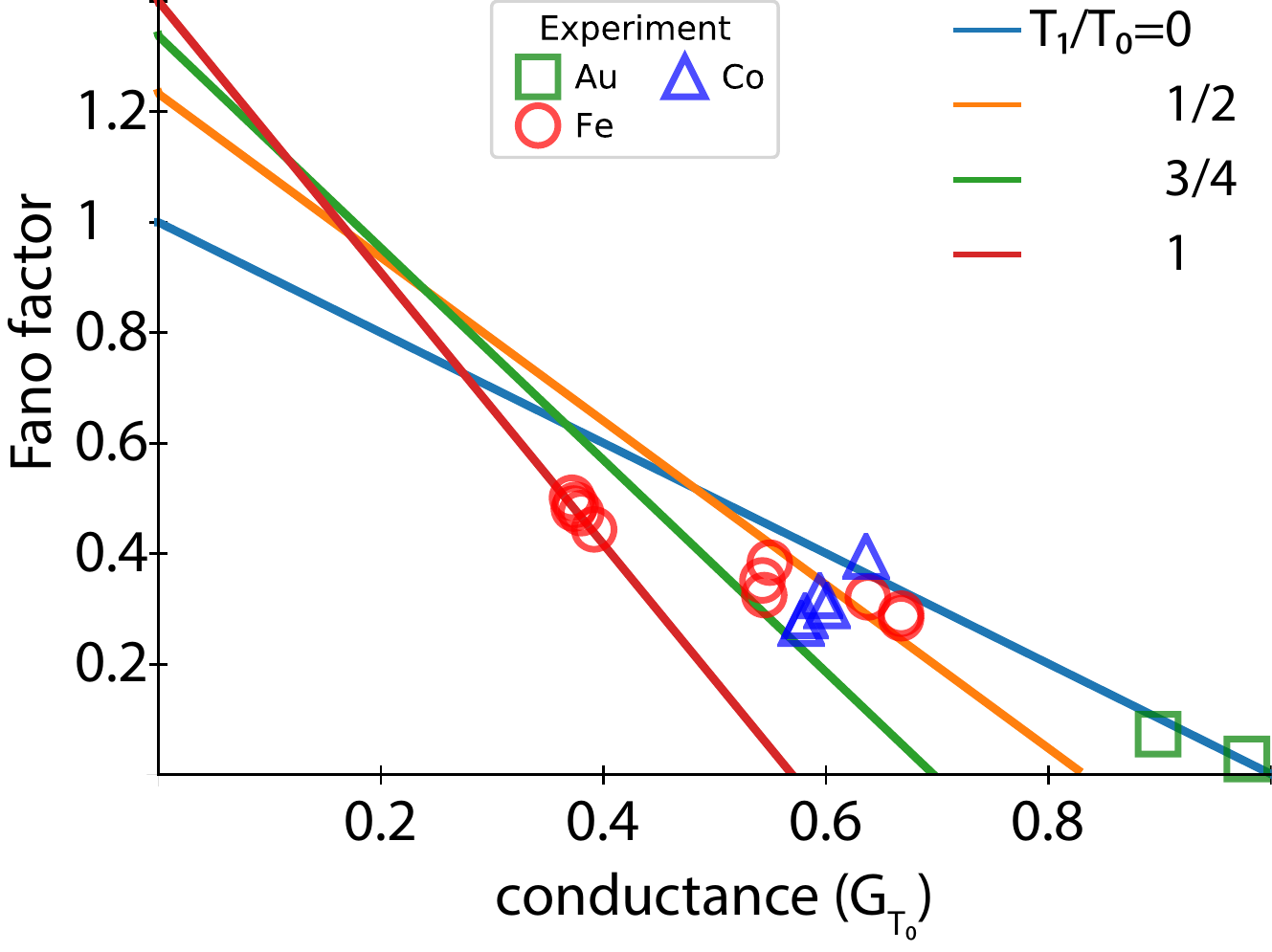}
\caption{(Color online) The Fano factor as function of $G_{T_0}$ for $T_1/T_0=0$, $1/2$, $1$, $6/5$, and $3/2$. Other parameters are as in Fig. \ref{fig:noise}. Experimental Fano factors of Fe ($\circ$), Co ($\triangle$), and Au ($\Box$) adatoms on Au(111) are taken from \cite{PhysRevLett.114.016602}.}
\label{fig:fano}
\end{figure}

\jonas{
Although external magnetic fields can be used to access more details about the spin excitation spectrum, the resolution is, as always, limited by the effective temperature of the local environment. This can be seen in Fig. \ref{fig:BTnoise} (c), (d), where we plot the voltage dependence of the shot noise and differential shot noise, respectively, for increasing temperatures $k_BT/|D|=0.5$, 1, 2, 3.5, and 5. The shot noise reveals essentially the same behavior with decreasing temperatures as with increasing magnetic field strengths, in the sense that the low voltage shot noise goes from the rounded $U$ shape towards a chevron like shape. Also, at low temperatures, kinks associated with spin-correlations in the shot noise become visible. These features are transferred into clear steps in the differential shot noise for low temperatures, while these are effectively smeared out by the thermal excitations at higher temperatures. 
}
The Fano factor $F=\lim_{V\rightarrow 0}S(V)/2e\av{I(V)}$ provides a measure of the character of the noise as a noise to current ratio.
In absence of the indirect tunneling $T_1$, the Fano factor reduces to simply $F_{|T_1=0}=1-T_0^2$, at low temperatures, as shown in Fig. \ref{fig:fano} ($T_1/T_0=0$ -- blue line), where the Fano factor is plotted as function of the equilibrium conductance $G_{T_0}=\lim_{V\rightarrow0}d\av{I(V)}/dV$, and different ratios $T_1/T_0$. The limit line $1-T_0^2$ is known as Poissonian noise and signifies the characteristics of ideal independent tunneling processes, observed in, e.g., atomic size metallic tunnel junctions \cite{brom,safonov}.

Correlated tunneling processes modify the tunneling properties such that the Fano factor deviates from the ideal Poissonian limit. Thus, whenever the $F<1-T_0^2$ ($F>1-T_0^2$) it is referred to as sub- (super-) Poissonian, and both sub- and super-Poissonian noise have been observed in experiments \cite{PhysRevLett.114.016602,safonov}.

Inclusion of the indirect tunneling processes $(T_1>0$) shows a dramatic change of the Fano factor from the Poissonian nature to a non-trivial dependence on both the ratio $T_1/T_0$ and $G_{T_0}$, Fig. \ref{fig:fano}. While super-Poissonian noise tends to be dominant for small $T_1/T_0\lesssim1/2$ for a larger range of $G_{T_0}$, the noise becomes increasingly super-Poissonian for increasing ratio $T_1/T_0$, however, over a smaller range of $G_{T_0}$.
Although the Fano factor decreases monotonically with the conductance, being super- (sub-) Poissonian as $G_{T_0}\rightarrow0$ ($G_{T_0}\rightarrow1$), the transition between the two phases depends on $T_1/T_0$. This feature indicates that the local spin fluctuations play an important role whenever the signal-to-noise ratio is small and that its influence on the transport properties weakens as the conductance grows. This is, however, expected from the point of view that the number $N$ of electrons involved in the tunneling is small at low conductances and that the signal-to-noise ratio depends on $N$ roughly like $1/\sqrt{N}$ \cite{PhysRevB.66.195416}.

In order to make a direct comparison with recent experiments, we have included the data of atomic Fe ($\circ$), Co ($\triangle$), and Au ($\Box$), taken from \cite{PhysRevLett.114.016602}. We find that the presence of the localized spin and its exchange interactions with a portion of the tunneling current provides a simple and natural explanation for the observed of sub-Poissonian noise. This picture is also consistent with other types transport, e.g., differential conductance and inelastic electron tunneling spectroscopy, measurements performed on similar set-ups.

\jonas{
The agreement between the experimentally and theoretically obtained Fano factors is made for ratios $T_1/T_0$ varying between 1/2 and 1. These values are reasonable both in comparison with measurements of the differential conductance and inelastic electron tunneling spectroscpy \cite{Science.312.1021,Science.317.1199,Nature.467.1084,PhysRevLett.106.037205,PhysRevLett.111.157204,Science.350.417,PhysRevLett.114.106807} as well as from theoretical estimates concerning both single electron and Coulomb assisted tunneling rate \cite{PhysRevB.64.153403,arXiv:1007.1238}.
}

\jonas{
\section{Discussion and summary}
\label{sec-discussion}
Our derivation the shot noise formula, intended to be applicable to set-ups with a magnetic moment embedded in the junction between metallic leads, is based on a few assumptions, in addition to the ones already mentioned alongside the derivation. Here, we discuss whether these assumptions and simplifications are justified. We have, for instance, omitted possible contributions to the shot noise emerging from Kondo correlations. There are, at least, two reasons why our approach may be considered as a sufficiently good approximation even without the inclusion of such effects. First, previous studies of Fe and Co on various metallic surfaces have concluded the local moment of Fe to be larger than 1/2 while Co may also acquire a spin moment of 1/2, see, e.g., \cite{Science.300.1130,Science.320.82,PhysRevLett.102.257203,PhysRevLett.106.037205,PhysRevLett.119.197002}. However, the experimental observations reported in \cite{PhysRevLett.114.016602} does not indicate any significant qualitative difference in the properties of the shot noise, which should be expected if the Kondo correlations were of integral importance. Second, even if correlated processes that are omitted here, like Kondo screening, do contribute in an non-negligible way, these would lead to an enhancement of the non-Poissonian characteristics of the shot noise, since it is exactly the correlated tunneling processes that create deviations of the shot noise from the Poissonian limit. Hence, despite possible presence of higher order correlation processes, we obtain a good agreement with experimental observations although we have only accounted for the simplest possible correlation processes involved in the exchange interactions between the tunneling electrons and local spin moment. Although it is beyond the scope of the present article, it would, nonetheless, be desirable to also consider the contribution of Kondo screening to the shot noise.
}

\jonas{
The shot noise formalism is here based on non-equilibrium Green functions technique and it is, certainly, relevant to also ask to what extent we mean by non-equilibrium. In general, there is no restriction introduced when applying the Keldysh technique, however, there are yet several other simplifications that have to be discussed. The spin is, for instance, considered in the atomic limit, which is only valid whenever the spin dynamics is hardly affected by the tunneling current.   This is motivated when the local exchange integral $T_1$ between the tunneling electrons and the spin is smaller than, for instance, the energy required for the local spin to make a transition to an excited state. This is, however, typically always the scenario in tunneling measurements made using STM, since the tunneling rate depends exponentially on the distance between the tip and the sample. We notice, nonetheless, that renormalization of the local spin due to the current flow generates a decreased life-time of the spin states which may give quantitative changes in both the current and shot noise and, while it is an issue beyond the scope to the article, remains an open question for the characteristics of the shot noise.
}

\jonas{
It should be noticed, finally, that the evidence presented here suggesting that shot noise reported in \cite{PhysRevLett.114.016602} is due to spin-correlations is quite circumstantial. Thus far, the comparison is made only through the Fano factor and agreement with merely one quantity may be obtained by means of various approaches. While one for instance may use the uncorrelated Landauer-B\"uttiker approach as was done in \cite{PhysRevLett.114.016602} or the approach presented in this article, there may be other mechanisms in the transport properties that yield the same Fano factor. The fact that we base our discussion on a model that successfully has been use to reproduce differential conductance and inelastic tunneling electron spectroscopy is in favor of our spin-correlated picture, since it creates a consistent framework of the different aspects of using tunneling transport in studies of local spin moments. In order to find the arguments for discrimination between different theories it is, however, necessary with more experiments that can be used for the construction of a sound description of the shot noise of local spin moments.
}

In summary we have presented a theoretical account of spin noise spectroscopy using charge transport measurements, in which the transport channels is partitioned into contributions with and without spin correlations. The channel without spin correlations constitute the usual non-interaction Landauer-B\"uttiker picture whereas the spin correlated channel emerges from the exchange coupling between the tunneling electrons and the localized spin moment. We have shown that our approach provides \jonas{a good} agreement with recent shot noise experiments using STM \cite{PhysRevLett.114.016602}. Simultaneously, being fully consistent with previous theoretical approaches to inelastic electron tunneling spectroscopy recorded on magnetic adatoms \cite{NanoLett.9.2414,PhysRevLett.102.256802,PhysRevLett.103.050801,PhysRevLett.103.176601}, as well as electron paramagnetic resonance \cite{PhysRevB.66.195416,AdvPhys.61.117,SciRep.6.25584} using STM. By means of our results we make the prediction that details of the spin excitation spectrum should be conceivable through finite voltage noise spectroscopy, which would open new routes for analyses of spin moments and anisotropy parameters.

\end{document}